\documentclass[prb,preprint,showpacs,preprintnumbers]{revtex4}
\usepackage{graphicx}
\usepackage{subfigure}
\usepackage{amssymb,amsmath}
\begin{document}
\bibliographystyle{apsrev}

\title{Ab initio approach of the hydrogen insertion effects 
       on the magnetic properties of $ {\bf ScFe_2} $}

\author{A.\ F.\  Al Alam,$^a$
        S.\ F.\ Matar,$^a$\footnote{Corresponding author: matar$@$icmcb-bordeaux.cnrs.fr}
        N.\ Ouaini,$^b$ and  
        M.\ Nakhl$^b$}
\affiliation{$^a$ICMCB, CNRS, Universit\'e Bordeaux 1, 
             87 avenue du Docteur Albert Schweitzer, 
             33608 Pessac Cedex, France, \\
             $^b$Universit\'e Saint-Esprit de Kaslik, 
             Facult\'e des Sciences, 
             B.P. Jounieh, Lebanon.}

\date{\today}

\pacs{07.55.Jg, 71.20.-b, 71.23} 


\begin{abstract}
The electronic and magnetic structures of $ {\rm ScFe_2} $ and of its dihydride 
$ {\rm ScFe_2H_2} $ are self-consistently calculated within the density functional 
theory (DFT) using the all electron augmented spherical wave (ASW) method with the 
local spin density approximation (LSDA) for treating effects of exchange and correlation. 
The results of the enhancement of the magnetization upon hydrogen insertion are 
assessed within an analysis of the chemical bonding properties from which we suggest 
that both hydrogen bond with iron and cell expansion effects play a role in the 
change of the magnitude of magnetization. In agreement with average experimental 
findings for both the intermetallic system and its dihydride, the calculated Fermi 
contact terms $H_{FC}$ of the $^{57}$Fe M\"ossbauer spectroscopy for hyperfine field, 
at the two iron sites, exhibit an original inversion for the order of magnitudes upon 
hydriding.
\end{abstract}
\maketitle

\section{Introduction}
Binary alloys belonging to the Laves family $ {\rm AB_2} $ (A= rare earth or actinide, 
B = transition metal) crystallize either in a face centered cubic fcc (C15) lattice or 
in a hexagonal lattice (C14); further, dihexagonal (C36) minority structure exists.\ 
\cite{Laves,Constitution} Within this family, $ {\rm ScFe_2} $ exhibits polymorphism 
and can exist in the three above crystalline states.\ \cite{Constitution} One of the 
basic aspects of the electronic properties of the pure $ {\rm AB_2} $ alloy systems 
is the identification of the origin of the magnetism which can be either due to the 
transition metal or induced by the A metal, depending on the chemical nature of the 
involved species.\ \cite{Boring,Konishi} Among others,\ \cite{Bodak,Ishida} Smit and 
Buschow have studied the synthesis of the $ {\rm ScFe_2} $ compound \cite{Buschow} 
reporting experimental measurements for the average magnetic moment of iron and the 
effective hyperfine field $H_{eff}$. On the other hand, the interaction of these 
intermetallic phases with hydrogen was investigated in a number of works.\ \cite{Buschow,
Burnasheva1,Semenenko,Niarcos,Burnasheva2} Besides the large potential applications 
of Laves phases hydrides in the field of solid state storage of hydrogen for energetics,\ 
\cite{Schlapbach,Zuttel} there is a basic interest in studying the magnetic structure 
and the electronic properties due to H insertion. At this level of investigation concerning 
$ {\rm ScFe_2} $, $^{57}$Fe M\"ossbauer spectroscopic works show an increase for the 
average magnitudes of the magnetization and the hyperfine field for iron upon hydriding 
without specifically assigning a role for each one of the two iron sites.\ \cite{Buschow} 
This leads to suggest an interplay between magnetovolume and chemical effects brought 
by the cell expansion when hydrogen is inserted. This original feature is addressed in 
this work. Further, a detailed atom-resolved study of the magnetism is provided and the 
nature of the non-rigid-band behavior within $ {\rm ScFe_2} $ and its dihydride are assessed.

\section{Crystal structures} 
\begin{figure}[htbp]
\includegraphics[width=0.8\columnwidth ]{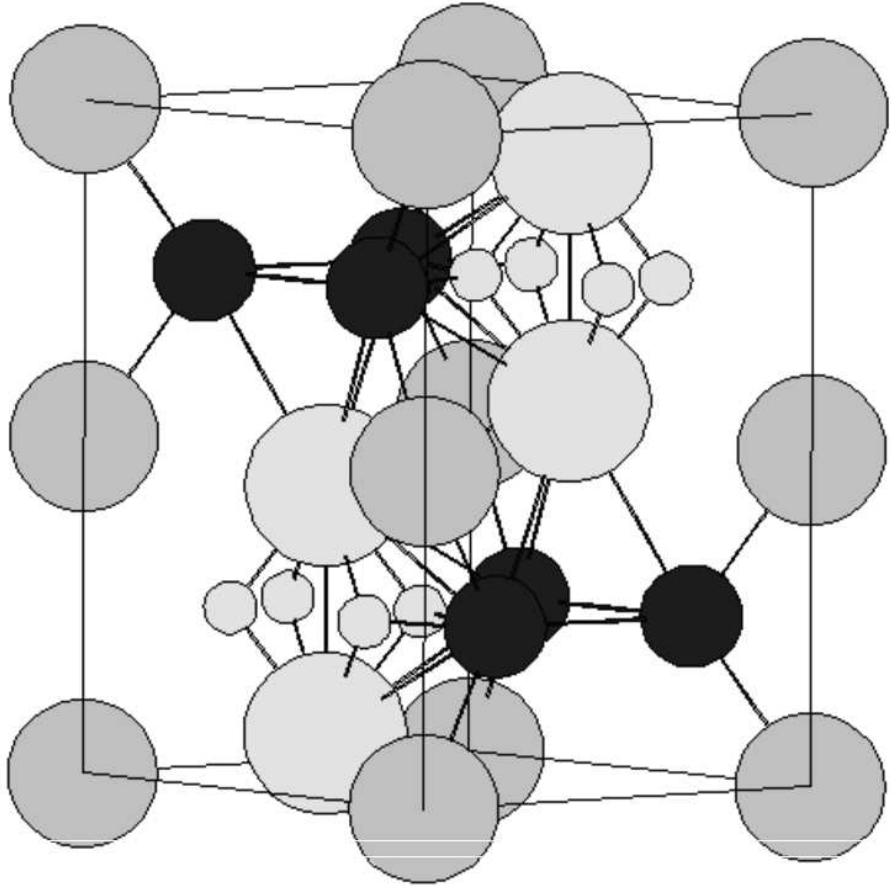}
\caption{The hexagonal crystal structure of $ {\rm ScFe_2H_2} $ ('C14' $Laves$ phase, space group $P6_3/mmc$). Sc (light grey), Fe1 (grey), Fe2 (black) and H (small light grey) are drawn in decreasing sphere sizes.}
		
\label{fig:struct}
\end{figure}

The $ {\rm ScFe_2} $ intermetallic system is experimentally stable in the hexagonal 
C14-type, $ P6_3/mmc $ space group, Laves structure.\ \cite{Ishida} For this alloy, 
Sc atoms are located in 4f sites at (1/3, 2/3, 0.0661). As for Fe atoms, there are 
two crystallographic nonequivalent sites: Fe1(2a) at (0, 0, 0) and Fe2(6h) at (0.8357, 
1.6714, 1/4). Fe2, Sc and Fe1 atoms have an occupancy ratio of 3 : 2 : 1. All sites 
mentioned in this report are in Wyckoff notation; numerical values are given in Refs.\ 
\cite{itc,Hong,Didisheim}. Smit and Buschow \cite{Buschow} charged an alloy sample with 
hydrogen gas, the composition of the formed hydride was found to correspond to $ \sim 1.92 $ 
atoms of H per formula unit (fu). Furtheron, this hydride will be referred to as $ {\rm ScFe_2H_2} $. 
A sketch of the structure containing hydrogen is presented in Fig.\ \ref{fig:struct}.

The choices for hydrogen insertion sites were done based on the neutron diffraction 
studies of Didisheim on the deuterated C14 compound $ {\rm ZrMn_2} $.\ \cite{Didisheim} 
Those studies show that hydrogen atoms in ternary hydrides, based on the C14-type, 
occupy interstitial tetrahedral-like $ A2B2 $ holes in positions such as: 24l, 12k, 
6h1 and 6h2.  Smit, Donkersloot and Buschow \cite{Smit}, have hydrided $ {\rm ScFe_2} $ 
into $ {\rm ScFe_2H_2} $ based on that same study, suggesting that each type of these 
holes is partially occupied by H atoms. In the present work, calculations were done 
with the assumption that hydrogens are located in sites such as: H1(6h1) at (0.463, 
0.926, 1/4) and H2(6h2) at (0.202, 0.404, 1/4). There are 12 of these tetrahedral holes 
per fu. Nevertheless, H atoms will occupy only 8 of these interstices with respect to 
the stoichiometry of $ {\rm Sc_4Fe_8H_8} $. The occupancy ratios of H1 and H2 atoms are 
the same as those of Fe2 and Fe1 in $ {\rm ScFe_2} $ respectively. Moreover, 12k and 24l 
interstices were tested. The performed calculations showed that they cannot reproduce 
magnitudes for the magnetic moments that are coherent with the expreriment. Values of 
the average magnetic moment for iron such as 1.865 and 2.065 $ \mu_B $, for 12k and 24l 
respectively, fail to compare with the experimental value of 2.23 $ \mu_B $.\ \cite{Buschow} 
These two $ A2B2 $ holes were discarded in favor of 6h1 and 6h2 intersitices which give 
more suitable values for the magnetic moments that agrees with experiment; these will be 
given later within the text. One can notice upon examining Fig.\ \ref{fig:struct} that Fe1 
atoms are little bonded with H atoms. As a matter of fact, $ A2B2 $ holes are formed by two 
$3d$ transition metal atoms (Fe1 and Fe2) and two atoms of the strongly hydrogen attracting 
component (Sc), which explains H atoms bonds with Sc. As for H-Fe2 bonds, table \ref{tab2} 
reports shorter Fe2-H distances compared to Sc-H one, which favors the H intake into $ A2B2 $ 
sites formed by Fe2 with respect to those formed by Fe1. Keeping in mind the relatively 
large difference in atomic volume between Sc and Fe, the choice of the H insertion sites 
is in agreement with the Westlake criterion (\cite{west3} and therein cited Refs.) that 
imposes a minimum interstitial hole size of 0.40 \AA . Too small H-H distances ($ d_{H-H} \geq  2.1 $\AA ) 
were avoided in order to respect the Switendick criterion.\ \cite{Switen} Table \ref{tab1} 
provides the crystal data of the two systems discussed in Ref.\ \cite{Buschow}. Note that 
there is a mismatch of the $ c/a $ ratio magnitudes for both experimentally prepared $ {\rm ScFe_2} $ 
and $ {\rm ScFe_2H_2} $ as stated in Ref.\ \cite{Buschow}, {\it i.e.}, $ c/a =1.636 $ for 
$ {\rm ScFe_2} $ and $ c/a=1.611 $ for $ {\rm ScFe_2H_2} $. Thus an anisotropic evolution 
accompanies the formation of the dihydride.

In this work hydrogen insertion effects within $ {\rm ScFe_2} $ are examined within three 
complementary approaches relevant (i) to the influence of volume expansion on the magnetic 
properties within the structure, (ii) to the chemical role of hydrogen and its influence on 
the changes of the magnetization for the dihydride lattice and (iii) to the crystal anisotropy 
occurring upon hydriding the alloy system. 

\begin{table}[htbp]
\begin{tabular}{lcccc}
\hline\hline
                &ScFe$_2$    &Expanded    &ScFe$_2$    &ScFe$_2$H$_2$ \\
\cline{3-4} 
                &$c/a$=1.636 &$c/a$=1.636 &$c/a$=1.611 &$c/a$=1.611 \\
\cline{2-2} \cline{3-4} \cline{5-5}  
a (\AA )        &4.963       &5.250       &5.277       &5.277 \\
c (\AA )        &8.122       &8.592       &8.504       &8.504 \\
Volume          &43.321      &51.276      &51.276      &51.276 \\
$ d_{Sc-Fe1} $  &2.915       &3.084       &3.100       &3.100 \\
                &            &3.110       &            & \\
$ d_{Sc-Fe2} $  &2.899       &3.052       &3.068       &3.068 \\
                &2.952       &3.084       &3.100       &3.100 \\
$ d_{Fe1-Fe2} $ &2.470       &2.592       &2.603       &2.603 \\
                &            &2.698       &            & \\
$ d_{H1-Sc} $   &            &            &            &1.963 \\
$ d_{H1-Fe2} $  &            &            &            &1.724 \\
$ d_{H2-Sc} $   &            &            &            &1.973 \\
$ d_{H2-Fe2} $  &            &            &            &1.698 \\
\hline\hline
\end{tabular}
\caption{$ {\rm ScFe_2} $ model systems: Crystal data for $ {\rm ScFe_2} $ 
         and $ {\rm ScFe_2H_2} $ \cite{Buschow} are presented in columns 
         1 and 4 respectively. Volume is given in ($ \AA^3/fu $ ). For more 
         details see text.}
\label{tab1}
\end{table}

\section{Computational framework}
\subsection{Electronic and magnetic properties}
Among the self-consistent methods built within the density functional theory (DFT) 
(see for instance \cite{dft1,dft2,dftmethods}) we use the augmented spherical wave 
(ASW) method \cite{asw1,asw2} which is an ``all-electrons'' method (non frozen core). 
The ASW method has proven its efficiency in similar topics of intermetallic systems 
and their hydrides.\ \cite{matar,chevalier} The analysis of the calculation results 
allows assigning a role to each atomic constituent in the magnetism and in the chemical 
bonding. The effects of exchange and correlation were treated within a local spin 
density approximation (LSDA) scheme.\ \cite{lsda} All valence electrons were treated 
as band states. In the minimal ASW basis set,\ \cite{asw2} we chose the outermost shells 
to represent the valence states and the matrix elements were constructed using partial 
waves up to $ l_{max.}+1=3 $ for Sc and Fe and $ l_{max.}+1=2 $ for H. The completeness 
of the valence basis set was checked for charge convergence meaning that charge residues are 
$\leq 0.1$ for $l_{max.}+1$. The self-consistent field calculations were run to a convergence 
of $ \Delta Q=10^{-8} $ for the charge density \cite{mixpap} and the accuracy of the method is 
in the range of about $ 10^{-7} $\,eV regarding energy differences. Besides its construction 
within the DFT, the ASW method is based on the atomic sphere approximation (ASA) which 
assumes overlapping spheres centered on the atomic sites within which the potential has 
a spherical symmetry -central potential-. The volume of the spheres has to be equal to 
the cell volume because the wave equation is solved only in the spheres. This is unproblematic 
for closely packed structures like metals and intermetallics such as $ {\rm ScFe_2} $ itself. 
But for less compacked structures such as that of $ {\rm ScFe_2H_2} $ studied here, additional 
augmentation spheres, called empty spheres (ES) are introduced to represent the interstitial 
space without loss of crystal symmetry and to avoid an otherwise too large overlap between 
the actual atomic spheres. ES are ``pseudo atoms'' with zero atomic number. They receive 
charges from the neighboring atomic species and allow for possible covalency effects within 
the lattice. Within the ASW method, the sphere geometry optimization (SGO) \cite{asw2} 
algorithm is used to generate ES without symmetry breaking. For $ \rm {ScFe_2H_2} $ one 
type of ES has been added. In total, 12 ES have been inserted into the $ \rm {Sc_4Fe_8H_8} $ 
structure. Besides obtaining the electronic band structure with site projected density of 
states (PDOS) for total spins (non spin polarized NSP configuration), as well as the spin-
resolved PDOS, the calculations allow discussing quantities such as the magnetic moments 
and their sign, the magnitude of the exchange splitting, as well as the spin densities at 
the core due to the $ns$ polarization thanks to the $d$ magnetic moment. These lead to the 
Fermi contact term of the hyperfine field $H_{FC}$ which constitutes the major part of the 
hyperfine field obtained by $^{57}$Fe M\"ossbauer spectroscopy. 

The calculations are started by assuming a non-magnetic configuration meaning a spin degeneracy for all valence states and equal spin occupations. Such a configuration should not be confused with a paramagnet, which could be simulated either by a supercell calculation with random spin orientations or by calling for disordered local moment approaches  
based on the coherent potential CPA approximation \cite{cpa} or the LDA+DMFT scheme.\cite{ldadmft} Subsequent spin-polarized calculations with different initial spin 
populations can lead at self-consistency either to finite or zero local 
moments within an implicit long-range ferromagnetic order.

\subsection{Chemical bonding properties}
The interactions within the alloy lattice with inserted hydrogen can be described 
in the framework of chemical bonding. Some elaborate tools exist allowing to obtain 
information about the nature of such interactions between atomic constituents as well 
as the respective quantum states involved. This can be provided using overlap population 
$ S_{ij} $ (OP) leading to the so-called crystal orbital overlap population (COOP) \cite{coop} 
or alternatively introducing the Hamiltonian based population ($ H_{ij} $) with the 
crystal orbital Hamiltonian population (COHP).\ \cite{cohp} Both approaches provide 
a qualitative description of the chemical interactions between two atomic species by 
assigning a bonding, non-bonding or anti-bonding character. A slight refinement of the 
COHP was recently proposed in form of the ``covalent bond energy'' ECOV which combines 
both COHP and COOP so as to make the resulting quantity independent of the choice of 
the zero of potential.\ \cite{ecov} Our experience with both COOP \cite{mat03} and ECOV 
\cite{mat07} shows that they give similar general trends although COOP exaggerate the 
magnitude of anti-bonding states. The ECOV criterion implemented within the ASW method 
is used here for the description of the chemical bond.  
  
\section{$ {\rm {\bf ScFe_2}} $ versus volume effects}
Calculations were performed at the experimental volume of $ {\rm ScFe_2} $.\ \cite{Buschow} 
In order to evaluate the magnetovolume effects, the computed results for the expanded 
hydrogen free $ {\rm ScFe_2} $ at the same lattice constants of the dihydride were addressed. 
As a matter of fact, such effects can be important in these intermetallic systems in as 
far as the onset of the magnetic moment is due to interband spin polarization, {\it i.e.}, 
it is mediated by the electron gas in a collective electrons approach. This is opposite to other systems, such as insulating oxydes where the magnetization is of intraband character, and hence, less affected by volume changes such as those induced by pressure (negative or positive).\ \cite{mat03}

\subsection{NSP calculations}
At self consistent convergence a progressive increase of BZ integration up to the value of 
$576$ {\bf k}-points, {\it i.e.}, $16$ {\bf k}-points in each direction of the irreducible 
wedge of the hexagonal BZ was used. A slight charge transfer of $ \sim 0.104 $ electron is 
seen from Fe2 towards Sc and Fe1. However its amount is not significant of an ionic behavior 
-rarely observed in the framework of {\it ab initio} calculations for such systems-\ \cite{Paul-Boncour}. 
Therefore it can be argued that the bonding is not mainly due to charge transfer but rather 
imposed by the hybridization between the different valence states. It is also important to 
mention that for all calculations (NSP as well as SP) the best evaluation of the radii for 
the different atomic species was assumed resulting in a better ASA overlap.

\begin{figure}[htbp]
\begin{center}
\includegraphics[width=\columnwidth]{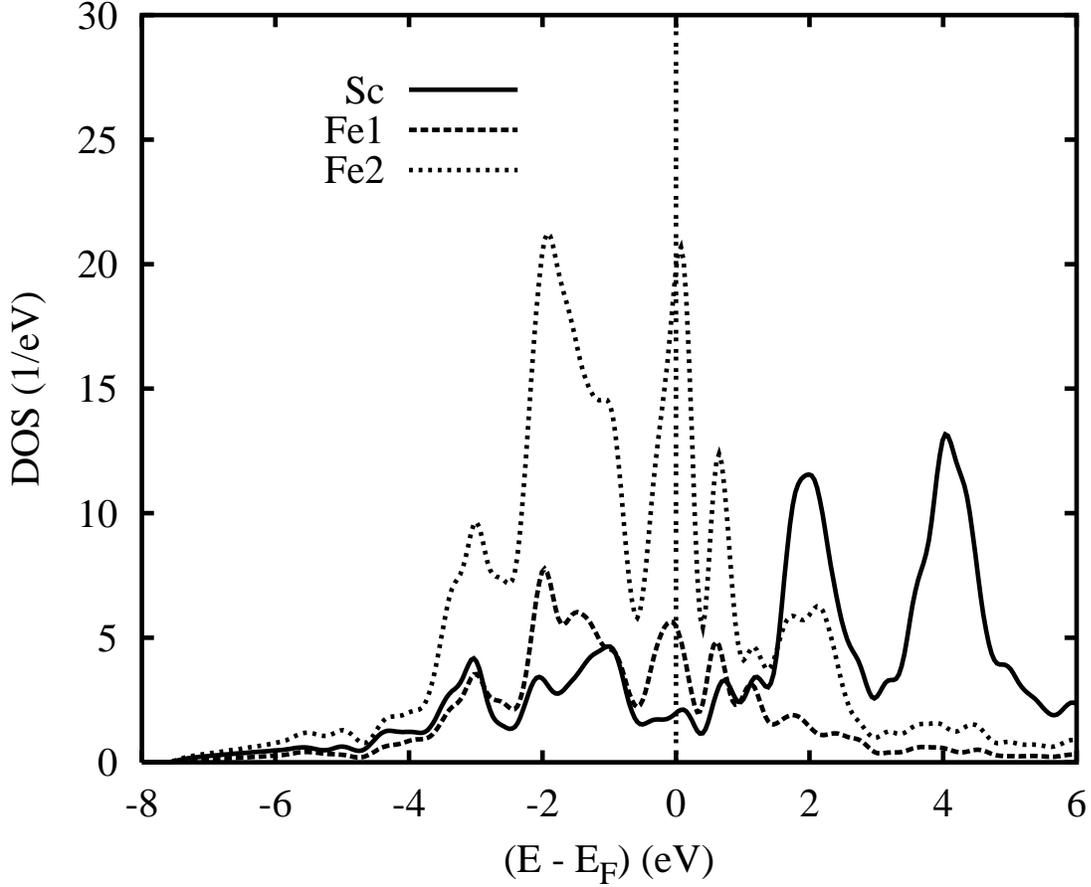}
\caption{Non magnetic site projected DOS of $ {\rm ScFe_2} $}
\label{DOS1}
\end{center}
\end{figure}
\subsubsection{Projected density of states PDOS} 
The PDOS for the $ {\rm ScFe_2} $ is given in Fig.\ \ref{DOS1} with respect to occupancy 
ratios given in section II (twice more Fe than Sc). This is applied for all the other PDOS 
pannels in this work. The origin of energies along the x axis is taken with respect to the 
Fermi energy ($E_F$); this is equally followed in all other plots. Looking firstly at the 
general shape of the PDOS one can observe that the $E_F$ level is situated at the peaks of 
both Fe1 and Fe2 with a predominance in terms of intensity for Fe2 states, {\it i.e.}, with 
respect to very low intensity scandium states. The similar skylines between the partial PDOS 
pointing to the mixing between Fe2, Fe1 and Sc states can be seen at the lower part of the 
valence band (VB), with mainly $sp$ like states between $-6$ and $-2.5$ \,eV, as well as 
towards the top of VB ($d$ states). Such mixing will be analyzed later regarding the chemical 
bonding. Lastly, within the conduction band (CB), $3d$(Sc) states are found dominant. This is 
expected as scandium is located at the very beginning of the $3d$ period, with mainly empty 
$d$ states.

\subsubsection{Analysis of the NSP-PDOS within the Stoner theory}
In as far as Fe1, Fe2 and Sc $3d$ states were treated as band states by our calculations, 
the Stoner theory of band ferromagnetism \cite{dftmethods} can be applied to address the spin 
polarization. The total energy of the spin system results from the exchange and kinetic energies 
counted from a non-magnetic state. Formulating the problem at zero temperature, one can express 
the total energy as $ E= \frac{1}{2} [\frac{m^2}{n(E_F)}][1- {\rm I} n(E_F)] $. Here ${\rm I}$ 
is the Stoner exchange-correlation integral which is an atomic quantity that can be derived from 
spin polarized calculations.\ \cite{Janak} $n(E_F)$ is the PDOS value for a given species at the 
Fermi Level in the non-magnetic state. The product ${\rm I}n(E_F)$ from the expression 
above provides a criterion for the stability of the spin system. The change from a non-magnetic configuration towards spin polarization is favorable when ${\rm I}n(E_F)>1$. The system then stabilizes through a gain of energy due to exchange. From Ref.\ \cite{Janak}, ${\rm I(Fe)}=0.4624$ \,eV and the computed $n(E_F)$ values for Fe1 and Fe2 are $ \sim 3.047 $ and $3.286$ \,eV$^{-1}$ respectively. The Stoner products for Fe1 and Fe2 are then $ \sim 1.409 $ and $ \sim 1.519 $ respectively. This means that the Stoner criterion is satisfied for the two iron sites, within $ {\rm ScFe_2} $.

\subsubsection{Covalent bond energy ECOV}
\begin{figure}[htbp]
\begin{center}
\includegraphics[width=\columnwidth]{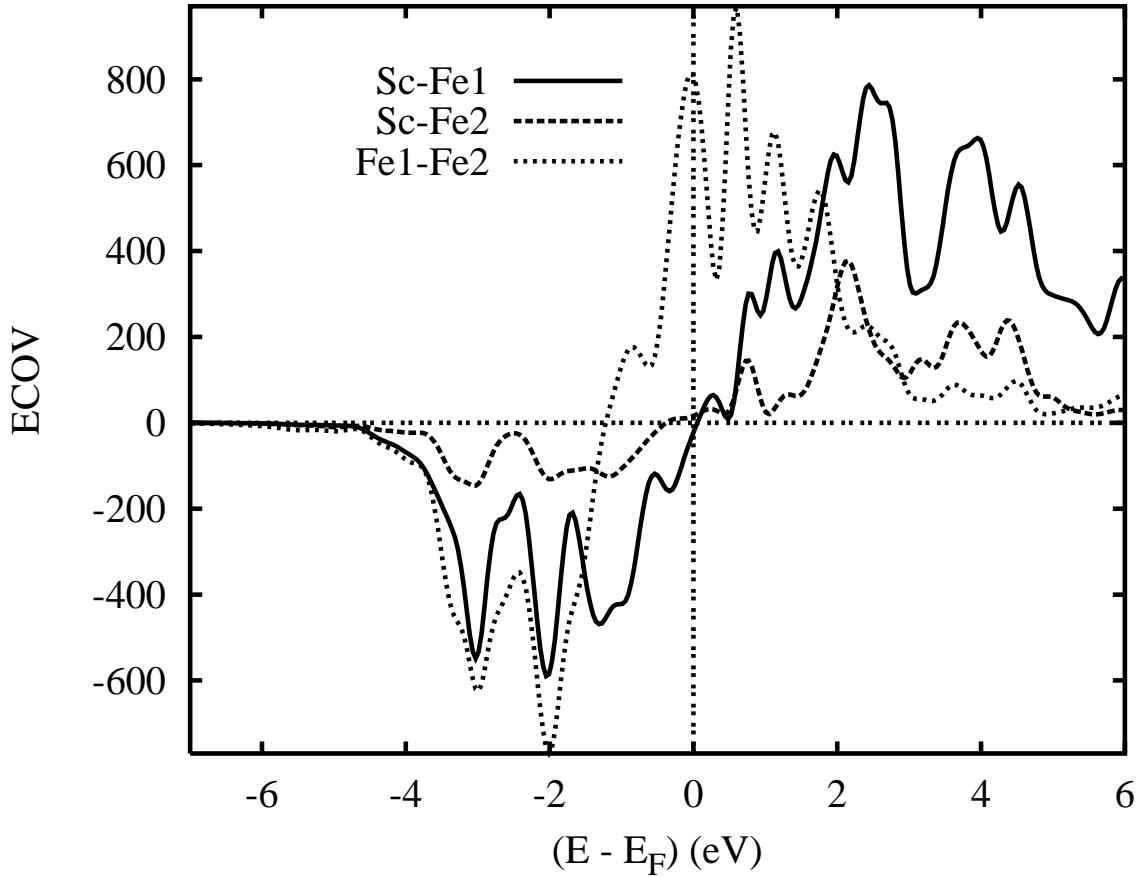}
\caption{Chemical bonding: non magnetic ECOV for $ {\rm ScFe_2} $.}
\label{ECOV1}
\end{center}
\end{figure}

The analysis of the chemical bonding is done using the ECOV approach presented in section III.B\ . 
The corresponding plots are shown in Fig.\ \ref{ECOV1}.

\begin{figure}[htbp]
\begin{center}
\subfigure{\includegraphics[width=\columnwidth]{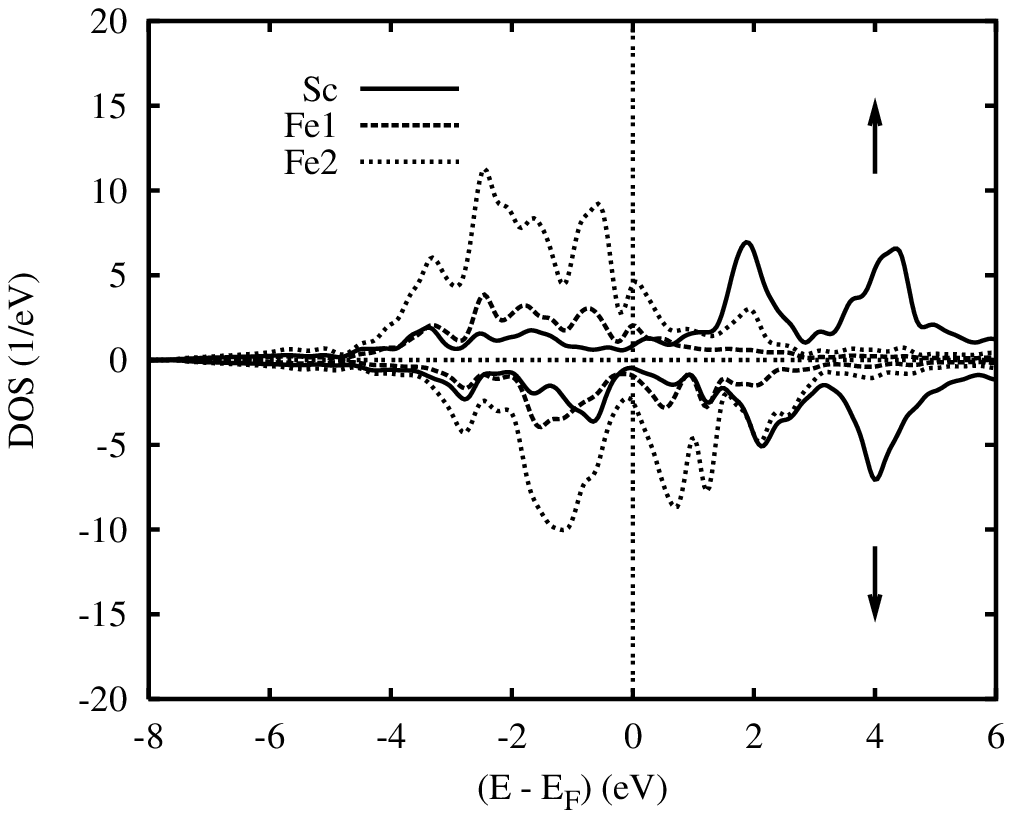}}
\subfigure{\includegraphics[width=\columnwidth]{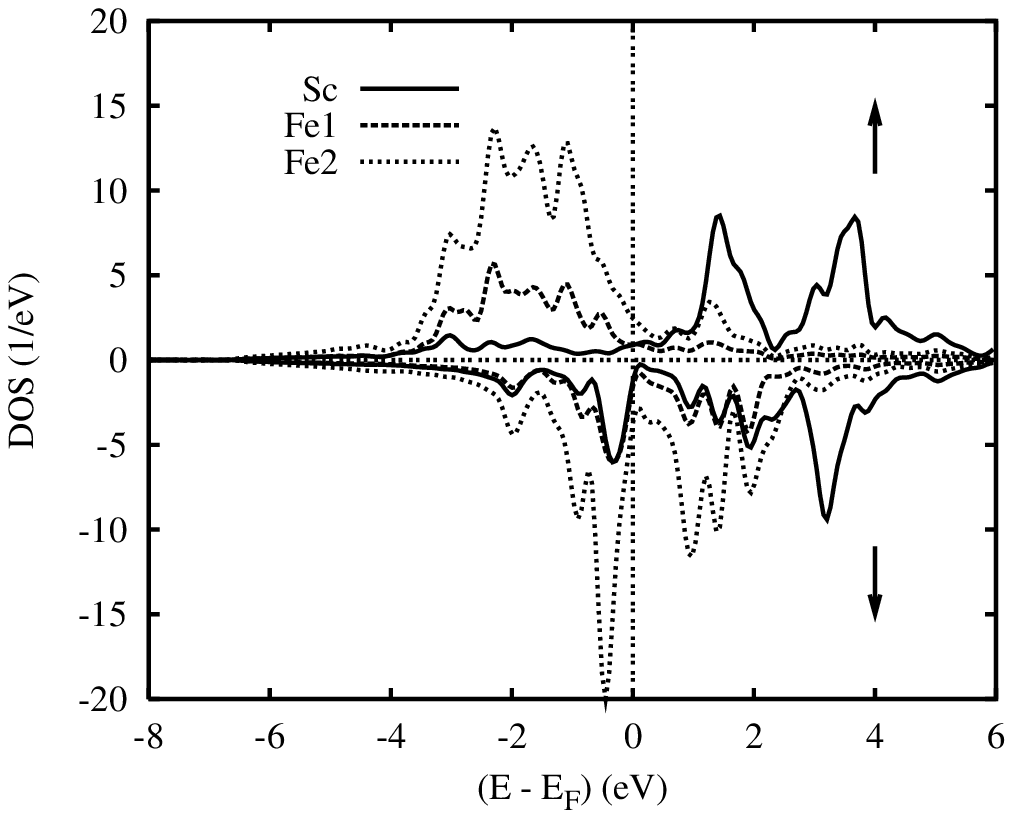}}
\caption{Site and spin projected DOS of $ {\rm ScFe_2} $ (a) 
         and the expanded hydrogen free $ {\rm ScFe_2} $ (b)}
\label{DOS2}
\end{center}
\end{figure}
\begin{figure}[htbp]
\begin{center}
\includegraphics[width=\columnwidth]{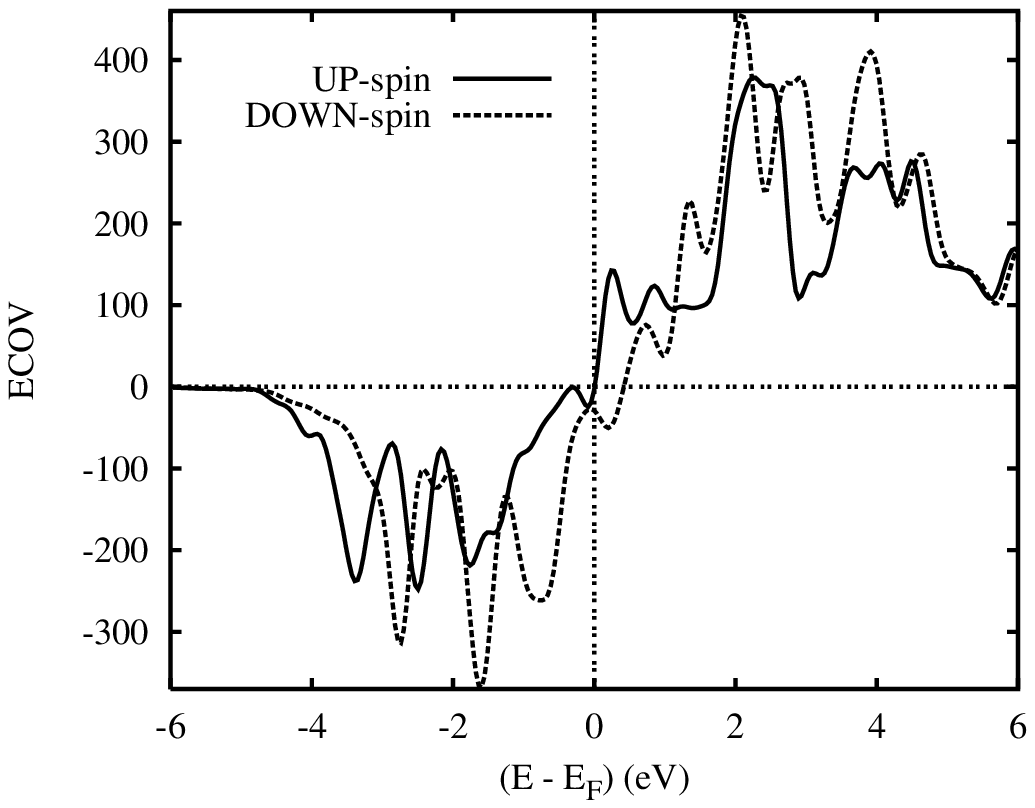}
\caption{Spin resolved chemical bonding: $ \uparrow $ and $ \downarrow$ ECOV-spin for Sc-Fe1 
         atomic pair interaction within $ {\rm ScFe_2} $.}
\label{ECOV2}
\end{center}
\end{figure}
Along the coordinate axis negative, positive and zero ECOV magnitudes (unitless) point to 
bonding, anti-bonding and non-bonding interactions respectively. Partial ECOV are given for 
the atomic pair interactions of Sc-Fe1, Sc-Fe2 and Fe1-Fe2 bonds. The other ECOV plots 
presented in this work equaly describe atomic-pair interactions. The change in bonding 
strength is proportional to the distance magnitudes given in table \ref{tab1}, {\it i.e.}, 
the shortest interatomic distances characterize the strongest interactions. The anti-bonding 
character of the strongest Fe1-Fe2 interaction at the top of the VB and at $E_F$ points to 
the instability of the system in the non-magnetic NSP configuration. On the contrary, the 
bonding character of the Sc-Fe1 interaction up to $E_F$ contributes to the stability of the 
system. The $d$ band electrons crossed by the Fermi level are not all anti-bonding. A part 
of those electrons becomes non-bonding in the neighborhood of $E_F$, thus participating to 
the onset of the magnetic moment.  

\subsection{Spin polarized calculations}
As it can be expected for the magnetic configuration of $ {\rm ScFe_2} $, there is an energy 
stabilization of $ \Delta E = 0.218 $ \,eV per fu with respect to the NSP calculations. This 
agrees with the experimental ferromagnetic ground state whereby $ {\rm ScFe_2} $ is considered 
as the only ferromagnet among C14 stoichiometric transition metals compounds as described in 
Refs.\ \cite{Ishida,Ikeda,Sankar}.

\begin{table}[htbp]
\begin{tabular}{lcccc}
\hline\hline
                      &ScFe$_2$    &Expanded    &ScFe$_2$    &ScFe$_2$H$_2$ \\
\cline{3-4}
                      &$c/a$=1.636 &$c/a$=1.636 &$c/a$=1.611 &$c/a$=1.611 \\
\cline{2-2} \cline{3-4} \cline{5-5}  
$m_{Sc}$              &-0.482      &-0.676      &-0.682      &-0.347 \\
$m_{Fe1}$             &1.468       &2.229       &2.279       &2.486 \\
$m_{Fe2}$             &1.560       &2.101       &2.118       &1.967 \\
$< m_{Fe} >$          &1.514       &2.165       &2.198       &2.226 \\
$M$                   &2.600       &3.591       &3.635       &3.791 \\
$H_{FC}^{total}(Fe1)$ &-157        &-240        &-246        &-203 \\
$H_{FC}^{total}(Fe2)$ &-162        &-229        &-226        &-232 \\
$H_{FC}^{core}(Fe1)$  &-159        &-252        &-255        &-272 \\
$H_{FC}^{core}(Fe2)$  &-164        &-232        &-231        &-214 \\
$E_{rel}$             &24.733      &26.661      &26.671      &0.000 \\ 
\hline\hline
\end{tabular}
\caption{Magnetic results for $ {\rm ScFe_2} $ calculated in this work.  
         Magnetic moments and total values are given in $ \mu_B $. 
         The core part $H_{FC}^{core}$ of the fermi contact term of $H_{eff}$ 
         as well as $H_{FC}^{total}$ which is the sum of both core and valence 
         contributions have their calculated value listed in \,kGauss. $E_{rel}$ 
         represents the relative SP-energy per fu towards the most stable value 
         of $ -89884.16355 $ \,eV corresponding to $ {\rm ScFe_2H_2} $. For more 
         details see text.}
\label{tab2}
\end{table}

\subsubsection{Magnetic moments}
Magnetic moments are obtained from the charge difference between $ \uparrow-spin $ 
and $ \downarrow-spin $ of all valence states; their calculated values are listed in 
table \ref{tab2}. The computed values for both the average magnetic moment of iron and 
the magnetization per fu are $< m_{Fe} > = 1.514 ~\mu_B$ and $M = 2.600 ~\mu_B$. Different 
experimental values were obtained by magnetic measurements; $M = 2.9 ~\mu_B$ \cite{Sankar} 
and $< m_{Fe} > = 1.2, 1.37, 1.45 ~\mu_B$ given in Refs.\ \cite{Nishihara,Ikeda,Sankar} 
respectively. Smit and Buschow also reported in Ref.\ \cite{Smit} that $< m_{Fe} >$ is 
equal to $ 1.14 ~\mu_B $ for sub-stoichiometric $ {\rm ScFe_{1.96}} $ and to $ 1.34 ~\mu_B $ 
for over-stoichiometric $ {\rm ScFe_{2.05}} $. It can be then suggested that the values 
in this work are within the range of the experimental data. Also the scandium carries a 
negatif magnetic moment of  $ -0.482 ~\mu_B $. From this $ {\rm ScFe_2} $ is a ferrimagnet 
in its ground state, contrary to some experimental results \cite{Buschow,Ishida} which 
announce it as a ferromagnet. The same ordering was observed for yttrium, within $ {\rm YFe_2} $, 
which carries a magnetic moment of $ -0.50 ~\mu_B $.\ \cite{Paul-Boncour} Moreover, 
the $3d$(Sc) orbital holds a calculated value of  $ -0.347 ~\mu_B $ which stands out as the 
largest contribution within the magnetic moment. The second largest contribution is that of 
the $4p$(Sc) orbital with a value of $ -0.103 ~\mu_B $. On the other hand, one can establish 
on analyzing the magnetic results for the expanded hydrogen free $ {\rm ScFe_2} $ given in 
table \ref{tab2} that $ {\rm ScFe_2} $ gives rise to an increase in the magnetic moments 
($ \Delta m / m $) for all the atomic species upon volume expansion. This increase is around 
$ 64 \% $ for the average magnetic moment of iron. Experimental values \cite{Smit} are found 
such as $ \Delta m / m = 0.96 $ and $ 0.37 $ for $ {\rm ScFe_{1.96}} $ and $ {\rm ScFe_{2.05}} $ 
respectively. The fact that the calculated value ranges between these two experimental values 
is related to the stoichiometric ordered lattice assumed by the calculations. Furthermore, 
the values $ \Delta m / m = 0.46, 0.75 $ and $0.87$ for Fe2, Sc and Fe1 respectively show 
that this increase is not at the same rate for the different species. This can be explained 
in terms of a reduction in contact between the scandium and iron atoms.\ \cite{Smit} The 
interatomic distances values in table \ref{tab1} confirm this explanation where it is found 
that the distance between the atomic pair Sc-Fe2 is smaller than the Sc-Fe1 distance. This is 
opposite to the $ {\rm ScFe_2} $ alloy system where $ d_{\rm Sc-Fe1} < d_{\rm Sc-Fe2} $ and 
$ m_{Fe1} < m_{Fe2} $. Lastly, table \ref{tab2} reports a higher $m_{Fe2}$ magnitude with 
respect to $m_{Fe1}$ for the intermetallic $ {\rm ScFe_2} $. This order of magnitudes is not 
respected for the expanded hydrogen free $ {\rm ScFe_2} $, where $m_{Fe2}$ is smaller than 
$m_{Fe1}$. The calculations show an increase in the difference between $ \uparrow $ and 
$ \downarrow-spin $ for the $d$ states upon volume expansion for both Fe1 and Fe2. This 
increase is around $85$ and $ 46 \% $ for Fe1 and Fe2 respectively. This major difference 
explains the inversion for the order of magnitudes brought by volume expansion.

\subsubsection{Hyperfine field $H_{FC}$}
Another significant result extracted from these calculations is the Fermi contact term 
of the hyperfine field ($H_{FC}$) (see table \ref{tab2}). The effective magnetic field 
$H_{eff}$ acting on a nucleus is considered as the sum of four contributions; (i) $H_i$, 
the internal field which is the magnetic field at the nucleus generated from an externally 
applied field, (ii) $H_{FC}$, the Fermi contact term, based on the spin density at the nucleus 
for the $ns$ quantum states caused by the polarization of the $s$ electrons by the $d$ 
moments, (iii) $H_{orb}$, which is the field arising from the orbital magnetic moment and 
(iv) $H_{dip}$, representing the dipole interaction with the surrounding atoms. In a 
non-relativistic description, $H_{FC}$ is expressed by the formula: 
$ H_{FC} = - \frac{8\pi}{3} \gamma_N \{(\Phi_{\uparrow}(0))^2-(\Phi_{\downarrow}(0))^2\} $. 
Where $\gamma_N$ is the nuclear gyromagnetic ratio and the quantities between brackets are 
the densities of $s$ electrons at the nucleus ($ r = 0 $) for $ \uparrow $ and $ \downarrow-spin $ 
respectively. The calculated $H_{FC}$ values are such as: $ H_{FC}^{total}(Fe1) = -157 $ 
\,kGauss and $ H_{FC}^{total}(Fe2) = -162$ \,kGauss. Based on $^{57}$Fe M\"ossbauer spectroscopy, 
reported experimental values of $H_{eff}$ \cite{Buschow} are $-167$ and $-174$ \,kGauss for 
Fe1 and Fe2 respectively. Other experimental values of the average hyperfine field were 
observed to be $-170$ \,kGauss for Fe1 and $-176$ \,kGauss for Fe2 for $ {\rm ScFe_{1.96}} $.\ 
\cite{Smit} Magnitudes of $-220$ and $-203$ \,kGauss, for the two crystallographic iron sites 
2a and 6h respectively, are found for $ {\rm ScFe_{2.05}} $ in Ref.\ \cite{Smit}. This is an 
experimental evidence of the inversion for the order of magnitudes of $H_{eff}$ in the 
intermetallic $ {\rm ScFe_2} $ due to stoichiometric changes. Such small departures from 
stoichiometry cannot be studied in the scheme of the calculations performed in this work; 
they require other schematic representations such as with the CPA.\ \cite{cpa} The difference 
between calculated and experimental values can be related to different origins relevant to 
(i) the fact that the local spin density approximation cannot treat with sufficient accuracy 
the polarization of core wave functions,\ \cite{Richter} (ii) the non-stoichiometry of the 
experimetally prepared alloys and the subsequent disorder within the solid solutions. An 
explanation for this peculiar behavior, {\it i.e.} the difference between calculated and 
measured values, can be found by decomposing $H_{FC}$ into its major contributions namely 
the one from the core $1s$, $2s$ and $3s$ electrons $H_{FC}^{core}$ and the one from the 
valence $4s$ electrons $H_{FC}^{val}$.\ \cite{mat99} While $H_{FC}^{core}$ usually is 
strictly proportional to the magnetic moment, $H_{FC}^{val}$ contains large contributions 
from the neighboring atoms. The calculated $H_{FC}^{total}$ values presented whitin this 
work are the sum of these two parts. Considering only the $H_{FC}^{core}$ contribution, 
values such as $-159$ and $-164$ \,kGauss were obtained for Fe1 and Fe2 respectively. 
This is a slight enhancement of the calculations with respect to experimental measurements.  

\subsubsection{Projected density of states PDOS}
The PDOS curves for the spin polarized SP configuration of $ {\rm ScFe_2} $ are shown 
in Fig.\ \ref{DOS2}(a). Within the VB two energy regions can be identified, from $-7.5$ 
to $-5$ \,eV, low intensity itinerant $s,p$ states of all constituents are found; this 
is followed by larger intensity peaks mainly due to $3d$(Fe) up to and above $E_F$. 
Exchange splitting can be seen to mainly affect the latter as it is expected from the 
above analysis of the magnetizations. $ \uparrow-spin $ states, for both Fe1 and Fe2 at 
$E_F$, are concentrated in sharp and narrow PDOS peaks, contrary to $ \downarrow-spin $ 
states that are found in PDOS minima. Fe1 and Fe2 peaks within the energy range 
$ [-1.3, -0.3] $ \,eV for $ \uparrow-spin $ states are similar to those 
corresponding to $ [0, 1] $ \,eV for $ \downarrow-spin $ states. This shift in 
spectral weight for $ \uparrow-spin $ states below the Fermi level and for $ \downarrow-spin $ 
states above $E_F$ corresponds to the onset of magnetic moments carried by Fe1 and Fe2. 
One can attribute this to a Stoner rigid-band magnetism at first sight. But PDOS weights 
at $ \uparrow $ and $ \downarrow-spin $ populations are not the same. This mismatch between 
both spin populations is mainly due to the $3d$(Fe2) states peaks at $\sim 1 $ \,eV for the 
$ \downarrow-spin $ states. This implies a non rigid-band shift which rules the magnetism 
of this system unlike $ \alpha $Fe.\ \cite{dftmethods} Magnetism arising in this way is 
called ``covalent magnetism''.\ \cite{williams} Such behavior was formerly shown for 
$ {\rm ZrFe_2} $.\ \cite{yfe2} One also notices that the peaks at $ \sim -0.7 $ \,eV 
for $ \downarrow-spin $ $3d$(Sc) states are more intense than those for $ \uparrow-spin $ 
states, a magnetic moment is carried by Sc with smaller magnitude and opposite direction 
to those carried by Fe1 and Fe2. The moment of scandium is provided by the covalent Sc-Fe1 
bond, rather than by a rigid energy shift of non-magnetic PDOS, whence its negative sign 
-notice the Sc-Fe1 overlap around $-0.7$ \,eV for $ \downarrow-spin $ PDOS -. On the other 
hand, the SP-PDOS for the expanded hydrogen free $ {\rm ScFe_2} $, given in Fig.\ \ref{DOS2}, 
show through the intense peaks at $ \sim -0.5 $ \,eV, a larger $ \downarrow - spin $ states 
occupation near $E_F$ in respect with the $ \uparrow - spin $ states. This feature, brought 
by volume expansion, is responsible of the inversion for the order of magnitudes of magnetic 
moments (see section IV.B.1). Another important feature is the Sc intense peak for $ \downarrow - spin $ at $ \sim -0.5 $ \,eV which is shifted closer to $E_F$ and is found more intense with respect to the scandium peak at $ \sim -0.7 $ \,eV in Fig.\ \ref{DOS2}(b). This peak is responsible for the increasing of the magnetic moment for Sc (see section IV.B.1) which is related to volume expansion.

\subsubsection{Spin resolved chemical bonding}
The Sc-Fe1 interactions for $ \downarrow $ and $ \uparrow-spin $ states curves are given in 
Fig.\ \ref{ECOV2}. The plots show that the $ \downarrow-spin $ bonding is stronger with respect to the $ \uparrow-spin $ one. This points to spin resolved chemical bonding. A peak at $ \sim -0.7 $ \,eV is pointing for the Sc-Fe1 $ \downarrow-spin $ interactions confirming this bond contribution to the magnetic moment carried by scandium.   

\section{$ {\rm {\bf ScFe_2}} $ and hydrogen effects}
For $ {\rm ScFe_2H_2} $, computed at the experimental data given in Ref.\ \cite{Buschow}, 
the BZ integration within the self-consistent calculations was carried out up to $1024$ {\bf k}
-points. Charge transfer within $ {\rm ScFe_2H_2} $ shows a departure of $ \sim 0.55 $ electron from Sc spheres to Fe1, Fe2 and empty spheres. This slight transfer, not significant of ionic effects, signals a redistribution of the two $s$ electrons of Sc over its three valence basis sets thus providing it with a larger $d$ character arising from its mixing with $3d$(Fe1) and $3d$(Fe2). The covalent bond between the metal species and H can be rather dicussed using the PDOS and ECOV criteria. The magnetic configuration of the dihydride is, as expected, energetically stabilized by $ \Delta E = 0.436 $ \,eV and $15.451$ \,eV per fu compared to the NSP calculations performed for $ {\rm ScFe_2H_2} $ and the SP calculations for $ {\rm ScFe_2} $ respectively. The relative energies per fu ($E_{rel}$) for the magnetic SP configuration, reported in table \ref{tab2}, describe the difference between the most stable $ {\rm ScFe_2H_2} $ energy and the energies of the other computed systems. The values for $E_{rel}$ show that while the volume expansion destabilizes the system, the hydriding restores it to an even more stable state. Using an equation giving the binding energy of the hydride system follwoing\ \cite{Hong} for 2H per fu: $E_B=\frac{1}{2} \left[E({\rm ScFe_2H_2})-E({\rm ScFe_2})-E({\rm H_2})\right]$, a calculation for the binding energy of the system due to  hydrogen is performed and the resulting value of $-0.92$ \,eV.mol$^{-1}$ is found in agreement with values for other systems.\cite{Hong} This shows the important role of hydrogen in the stabilization of the intermetallic system.  

\subsection{Magnetic moments}
Magnetic moments values derived from spin polarized calculations are given in table \ref{tab2}. The average magnetic moment for iron is of $ 2.226 ~\mu_B $ which agrees with the experimental result of $ 2.23 ~\mu_B $ given in Ref.\ \cite{Buschow}. This computed value corresponds to the increase $ \Delta m / m = 0.67 $, which is an enhancement of $ 1.3 \% $ over the increase brought by the volume expansion. Moreover, the respective $ \Delta m / m $ values for metal species (with respect to $ {\rm ScFe_2} $) show a decrease of $ 11 \% $ for Sc as well as an increase of $35$ and $ 104 \% $ for Fe2 and Fe1 respectively. This can be explained by an arrangement of the H atoms in the tetrahedral sites surrounding the Sc atoms, thus shielding the Fe1 sublattice from the Sc sublattice, as already suggested by Smit et al.\ \cite{Smit}, meaning a reduction in terms of contact between Sc and Fe1. Regarding the already mentioned role of the Sc-Fe1 interaction (see section IV.A.3) in the onset of the magnetic moment on scandium, this reduction is concomitant with a more pronounced magnetic moment for Fe1 and a less large one for Sc -in agreement with the computations-. The magnetic moment values given in table \ref{tab2} for Fe1 and Fe2 are $2.486$ and $1.967$ \,eV respectively. This inversion for the order of magnitudes, was already observed and explained for the expanded hydrogen free $ {\rm ScFe_2} $ (see section IV.B.1).   

\subsection{Hyperfine field}
The Fermi contact term of the hyperfine field calculated values for the dihydride 
are such as: $ H_{FC}^{total}(Fe1) = -203 $ \,kGauss and $ H_{FC}^{total}(Fe2) = -232 $ \,kGauss. The $H_{FC}^{core}$ computed values are such as $-272$ and $-214$ \,kGauss for Fe1 
and Fe2 respectively. Experimental findings in Ref.\ \cite{Buschow} of $-239$ and 
$-300$ \,kGauss correspond to $H_{eff}$. These values were given without being 
assigned to the two crystallographic sites (6h and 2a) for iron. Since the Fe1 
sublattice is shielded by hydrogen from the Sc sublattice, the contact between 
Fe1 atoms and the neighboring atomic species is reduced, consequently it can be assumed 
that $H_{FC}^{core}$ contribution is the one measured by the experiment for Fe1. 
It follows that the largest experimental value of -300 \,kGauss for $H_{eff}$ can 
be attributed to Fe1 for which the calculations have given a $H_{FC}^{core}$ equal 
to $-272$ \,kGauss. On the contrary, Fe2 atoms interactions with neighboring species 
are more important, hence the measured hyperfine field corresponds to the sum of both 
core and valence contributions of $H_{FC}$. Then the experimental value of $-239$ \,kGauss 
is assumed for Fe2, in agreement with the computed value of $-232$ \,kGauss for 
$H_{FC}^{total}(Fe2)$. On the other hand, the orders of magnitudes for $H_{FC}$ 
(see table \ref{tab2}) show that the highest value of $H_{FC}$ corresponds to the 
larger $m_{Fe}$ magnitude. Considering the core part for Fe1 on the one hand and the total contribution for $H_{FC}$ of Fe2 on the other hand, we can propose that this tendency is preserved upon hydriding. 

\subsection{Projected density of states PDOS}
\begin{figure}[htbp]
\begin{center}
\includegraphics[width=\columnwidth]{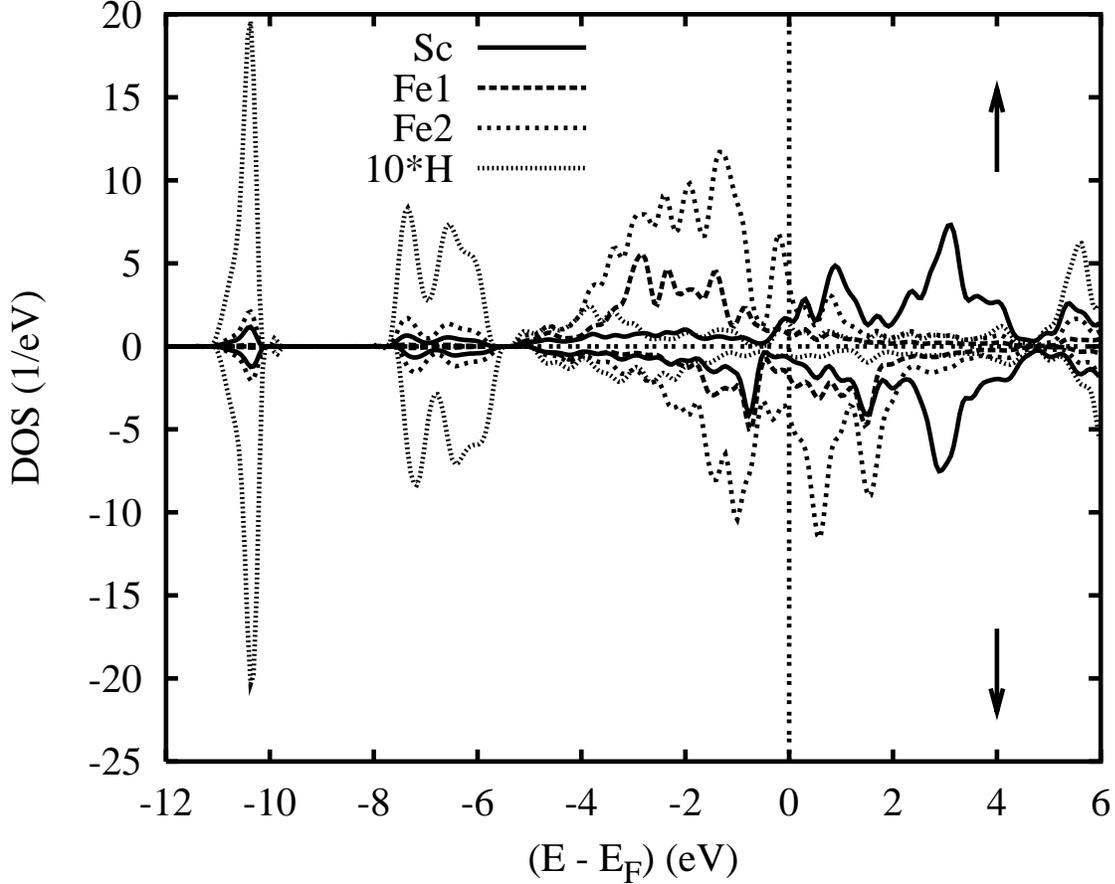}
\caption{Site and spin projected PDOS of $ {\rm ScFe_2H_2} $. For the sake of clear 
         presentation, hydrogen contribution was multiplied by 10.}
\label{DOS3}
\end{center}
\end{figure}

A visual inspection of the spin projected PDOS of $ {\rm ScFe_2H_2} $ represented in 
Fig.\ \ref{DOS3} shows two small energetic intervals ranging from $-8$ to $-5.5$ \,eV 
and from $-11$ to $-10$ \,eV respectively. The similar skylines between the different 
atomic species featuring in these two energy regions describe the hybridization of the 
metallic species with hydrogen. The sharper and narrower nature of these PDOS peaks, 
compared to those of $ {\rm ScFe_2} $ [Fig.\ \ref{DOS2}(a)] and the expanded $ {\rm ScFe_2} $ 
[Fig.\ \ref{DOS2}(b)], point to a larger localization of the states in the dihydride 
system. This is due to hydrogen intake and not to volume expansion.  

\begin{figure}[htbp]
\begin{center}
\includegraphics[width=\columnwidth]{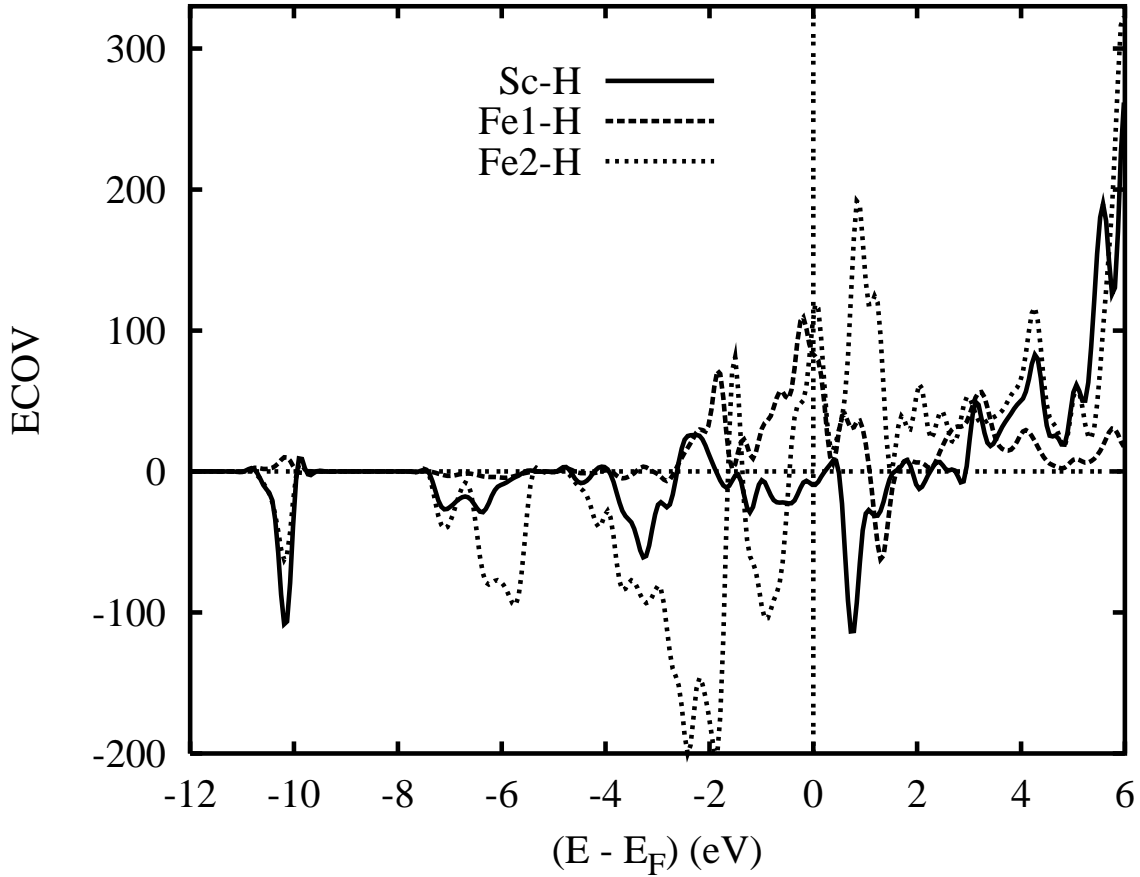}
\caption{Chemical bonding: non-magnetic ECOV within $ {\rm ScFe_2H_2} $ for the 
         atomic pair interactions Fe1-H, Fe2-H and Sc-H.}
\label{ECOV3}
\end{center}
\end{figure} 

\subsection{Chemical bonding}
To find an explanation for the magnetic behavior, chemical bonding plots with ECOV 
criterion (shown in fig.\ \ref{ECOV3}) for hydrogen interactions with the metal species 
from which the lattice is built are analyzed. In as far as $H_{FC}$ originates mainly 
from the spin polarization of valence states via interactions with the magnetic neighbors, 
the inversion for the order of magnitudes mentioned earlier for $H_{FC}$ is expained by 
the fact that both hydrogen insertion and shielding (see section V.A) reduce Fe1 interaction 
with neighbors. Fe1 atoms become further isolated upon hydriding, and tend towards 
magnetic behaviors of the same kind as in Fe metal. In this so-called weak ferromagnetism 
the majority spin subband as well as the minority spin subband are partially depleted.\ 
\cite{Friedel} This is plotted in Fig.\ \ref{ECOV3} where the Fe1-H interaction is 
anti-bonding through all the VB. This anti-bonding behavior confirms this reduction in 
contact between Fe1 and the other atomic species. 

\section{$ {\rm {\bf ScFe_2}} $ and anisotropy changes}
For the sake of addressing anisotropy effects, an additional expanded hydrogen free 
$ {\rm ScFe_2} $ model was calculated with the experimental value of the $c/a$ ratio 
of the alloy system.\ \cite{Buschow} The computed magnitudes and signs for the magnetic 
moments as well as for $H_{FC}$ are reported in table \ref{tab2}. An analysis of these 
results demonstrates that the change of the $c/a$ ratio does not affect the general 
trends of the magnetic behavior. Moreover, the already observed inversion (upon volume 
expansion) for the order of magnitudes for both magnetic moments and $H_{FC}$ is conserved. 
It can be then concluded that the anisotropy changes are of negligible influence on the 
magnetic behavior of $ {\rm ScFe_2H_2} $, {\it i.e.}, with respect to both volume expansion 
and hydrogen insertion effects.

\section{Conclusions}
In this work local spin density functional (LSDF) investigations of the hydrogen insertion 
effects on the magnetism and bonding within the C14 $ {\rm ScFe_2} $ laves phase have been 
undertaken. In order to address these features, we performed {\it ab initio} all electrons computations of the electronic band structure and of the bonding properties for $ {\rm ScFe_2} $ and its dihydride $ {\rm ScFe_2H_2}$ as well as for the expanded hydrogen free $ {\rm ScFe_2} $ system at the dihydride volume. Contrary to former studies \cite{Ishida,Ikeda,Sankar} which described the magnetic behavior of ScFe$_2$ by means of a rigid-band shift, our results point to a ``covalent magnetism'' -like behavior. The analysis of the electronic structures and of the chemical bonding properties using the covalent bond energy (ECOV) criterion leads to suggest that the volume expansion increases the magnitudes for both magnetic moments and $H_{FC}$. As for the chemical effect of hydriding, it further enhances the magnitude of the total magnetic moment, while the magnetic moments of the different atomic species show different behaviors. Furthermore, $H_{FC}$ values exhibit  orders of magnitudes which are proportional to those of the magnetic moments of Fe1 and Fe2 within $ {\rm ScFe_2} $, {\it i.e.}, both $m( Fe2)$ and $H_{FC}^{total}(Fe2)$ are larger than $m(Fe1)$ and $H_{FC}^{total}(Fe1)$. Volume expansion introduces an inversion for this order of magnitudes but respects the proportionality. However, upon hydriding, one should consider the core and valence contributions to $H_{FC}$. As a matter of fact, inserted H atoms shield the Fe1 sublattice, thus reducing contact with other sublattices. 
This leads to experimentally measured values of the hyperfine field for Fe1 concomitant with the calculated $H_{FC}^{core}$ which is proportional to the magnetic moment of the onsite $d$-electrons. On the other hand, in as far as the hydriding process introduces an anisotropic expansion to the $ {\rm ScFe_2} $ unit cell, further computations were performed for an expanded hydrogen free $ {\rm ScFe_2} $ model at the experimental $c/a$ ratio of $ {\rm ScFe_2} $ alloy system. The analysis of the results of these calculations helped showing that the anisotropy changes have no significant effects on the general magnetic behavior. 

\begin{acknowledgments}
Computational facilities were provided within the intensive numerical simulation facilities 
network M3PEC of the University Bordeaux 1, partly financed by the Conseil R\'egional d'Aquitaine.
\end{acknowledgments}


\end{document}